\documentclass[preprint,aps,prb,showpacs]{revtex4}
\usepackage{bm}
\usepackage{epsfig}
\usepackage{colordvi}
\newcommand{\nix}[1]{}

\usepackage{graphicx}
\begin{document}

\title{Magneto-gyrotropic effects in semiconductor quantum wells}
\author{V V~Bel'kov$^{1,2}$  and S~D~Ganichev$^{1}$
}
\address{$^1$ Terahertz Center, University of Regensburg,
93040 Regensburg, Germany}
\address{$^2$ A.F.~Ioffe Physico-Technical Institute, Russian
Academy of Sciences, 194021 St.~Petersburg, Russia}
%\ead{sergey.ganichev@physik.uni-regensburg.de}

\date{\today}

\begin{abstract}

Magneto-gyrotropic photogalvanic effects  in quantum wells are reviewed. 
We discuss experimental data, results of phenomenological analysis and microscopic models of these effects. 
The current flow is driven by spin-dependent scattering in low-dimensional 
structures gyrotropic media resulted in asymmetry of photoexcitation and relaxation processes. 
Several applications of the effects are also considered.

\end{abstract}


% 73.50.Pz Photoconduction and photovoltaic effects
% 72.25.Fe Optical creation of spin polarized carriers
% 72.25.Rb Spin relaxation and scattering
% 78.67.De Quantum wells

\maketitle

%\submitto{\SST}

\section{Introduction}

The spin of electrons and holes in solid state systems is an
intensively studied quantum mechanical property showing a large
variety of interesting physical phenomena. One of the most
frequently used and powerful methods of generation and
investigation of spin polarization is optical
orientation~\cite{Meier}. Besides purely optical phenomena, like
circularly polarized photoluminescence, the optical generation of
an unbalanced spin distribution in a semiconductor may lead to
spin photocurrents. Light propagating through a semiconductor and acting upon mobile
carriers can generate a $dc$ electric current, under short-circuit
condition, or a voltage, in the case of open-circuit samples. 
A spin photocurrent was proposed for the first time
in~\cite{Dyakonov71p144} (see also~\cite{Averkiev83p393}) and
thereafter observed in bulk AlGaAs~\cite{Bakun84p1293}. In these
works  it have been shown that an inhomogeneity of the 
%spin polarization of electrons 
spin polarized electrons
results in a surface current due to spin-orbit
interaction. A gradient of spin density was obtained by making use
of the strong fundamental absorption of circularly polarized light
at the band edge of the bulk semiconductor. Recent studies~\cite{Ganichev2006} demonstrated that 
spin photocurrents can also be generated by homogeneous spin 
polarization caused by  absorption of circularly polarized radiation
in low-dimensional systems, like circular photogalvanic effect
or spin-galvanic effect (SGE), or even as a result of the illumination 
with unpolarized radiation due to magneto-gyrotropic photogalvanic effects (MGE). 
All these phenomena are gathered in the class of photogalvanic effects 
(PGE), which, by definition, appear neither due to inhomogeneity of 
optical excitation of electron-hole pairs nor due to inhomogeneity 
of the sample.

In this paper we consider only the magnetic field induced photogalvanic effects in 
low-dimensional semiconductor structures. Moreover, we focus the attention here on the spin-photogalvanics
and discuss spin-related mechanisms of magneto-gyrotropic photogalvanic effects 
due to the Larmor precession induced spin-galvanic effect
and caused by zero-bias spin separation.
Microscopic mechanisms of these phenomena 
are given in section~\ref{II}. They are based on the spin-orbit coupling which 
provides a versatile tool to generate and to manipulate the spin degree of freedom 
in low-dimensional semiconductor structures. 
In general zero-bias spin separation and spin-galvanic effect
do not require an application of an external magnetic effect and for some mechanisms even light. 
However, they have been demonstrated and are most intensively studied applying MGE technique.
The macroscopic features of all magneto-gyrotropic effects discussed 
here, e.g., the possibility to generate a photocurrent, its behaviour upon variation of
radiation polarization, crystallographic orientation, experimental
geometry etc. 
are described in the frame of a phenomenological presented in section~\ref{III}.
Phenomenological theory  operates with conventional vectors, or {\it polar}
vectors, and pseudo-vectors, or {\it axial} vectors,  
and indeed does not depend on details of microscopic mechanisms.
section~\ref{IV} gives a short account of the experimental technique. 
In section~\ref{V} the experimental results are
presented and discussed in view of the theoretical background.
In this section we also discuss applications of MGE, in particular, we %focus on the
analyze spin splitting of subbands in  $\bm{k}$-space due to bulk inversion
asymmetry (BIA) and structural inversion asymmetry (SIA) in QWs of various
crystallographic orientations.

\section{Microscopic models}
\label{II}

\subsection{Spin-galvanic effect}

In a system of free carriers with nonequilibrium 
spin-state occupation but equilibrium energy distribution within
each spin branch, the spin relaxation can result in the generation of an electric current~\cite{Ivchenko89p175}.
%????? This effect, predicted by Ivchenko et al.\cite{Ivchenko89p175},
%was observed by Ganichev et al. applying THz radiation and named the spin-galvanic effect~\cite{Nature02}. 
This effect is called the spin-galvanic effect~\cite{Nature02}.
Phenomenologically, an electric
current can be linked to the electron's averaged spin 
%polarization {\boldmath$S$} by
polarization $\bm{S}$ by
\begin{equation} %%  1
j_{\alpha} =
\sum_{\gamma}Q_{\alpha \gamma} S_{\gamma}
\label{equ22}.
\end{equation}
For (001)-grown zinc-blende structure based  QWs of C$_{2v}$-symmetry  
this equation reduces to $j_{x} = Q_{xy}S_y$ and $j_{y} = Q_{yx}S_x$. 
Here we choose the coordinate system as 
$x \parallel [1 \bar{1} 0]$, $y \parallel [110]$ and $z \parallel [001]$.

Spin-galvanic effect generally does not require 
optical excitation, in fact, the nonequilibrium spin %$\bm{S}$ 
can be achieved both by
optical and non-optical methods, e.g., by electrical spin
injection. If the nonequilibrium spin, however, is produced by optical
orientation proportional to the degree of light circular 
polarization, $P_{circ}$, the current generation can be reputed  as 
a photogalvanic effect. Under pure optical excitation 
the spin-galvanic effect is usually   observed simultaneously with the circular 
photogalvanic effect. At this condition 
two effects can be experimentally separated in time resolved %measurements 
experiments or  by spectral measurements. However, an
another method, which, on the one hand, provides a 
nonequal population of
%uniform distribution in 
spin subbands and, on the other hand,
excludes the circular PGE was proposed in~\cite{Nature02}.

\begin{figure}[bt]%%  Fig.1
%\centerline{\psfig{file=spinfig07larmor.eps,width=3.65in}}
%\centerline{\epsfxsize 45mm\epsfbox{spinfig07larmor.eps}}
\centering{\epsfxsize 47mm\epsfbox{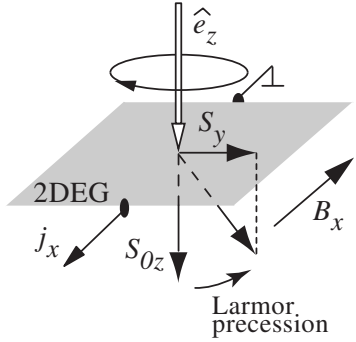}}   %%figure1 spinfig07larmor
\vspace*{8pt}
\caption{Optical scheme of generation a uniform  in-plane spin
polarization 
driving
%which causes 
a spin-galvanic current. Electron spins
are oriented normal to the QW plane  by circularly polarized
radiation and rotated into the plane by Larmor precession in an
in-plane magnetic field $B_x$ (after Ref.~\protect \cite{Nature02}).}
\label{spinfig07larmor}
\end{figure}

The method is based on the use of optical orientation 
at normal incidence and the assistance of an
external magnetic field to achieve an in-plane polarization in
(001)-grown low-dimensional structures (see figure~\ref{spinfig07larmor}). %~\cite{Nature02}. 
%The mostly used method to investigate the spin-galvanic effect is
%based on the irradiation of QWs by circularly polarized and application of the 
%in-plain magnetic field~\cite{Nature02}. 
For normal incidence the spin-galvanic effect 
as well as the circular PGE vanish. Thus, the spin %orientation 
polarization $S_{0z}$ along the $z$-axis is achieved but no
spin-induced photocurrent is generated~\cite{footnoteBakun}. 
An in-plane spin component, necessary for the spin-galvanic
effect, arises in an in-plane magnetic field.
The field perpendicular to the initially oriented spins 
(e.g., $\bm{B}\,\parallel x$) 
rotates
them into the plane of the two-dimensional electron gas (2DEG)  due to the Larmor
precession (Hanle effect). The nonequilibrium
spin polarization $S_y$ is given by
\begin{equation}  %%  2
S_y = - \frac{\omega_L \tau_{s\perp} }{1 + (\omega_L \tau_s )^2}\:
S_{0z}\:,
\label{Hanle}
\end{equation}
where $\tau_s = \sqrt{\tau_{s \parallel} \tau_{s \perp} }$,
$\tau_{s\parallel}$ and $\tau_{s\perp}$ are the
longitudinal and transverse electron spin relaxation times, and
$\omega_L$ is the Larmor frequency. Spin-galvanic effect investigated 
in such geometry belongs to the class of magneto-gyrotropic effects.
A characteristic feature of the magneto-gyrotropic effect due to the 
spin-galvanic effect is that the current  
reverses its direction upon changing the radiation helicity 
from left-handed to right-handed and vice versa as well as 
upon reversing of the in-plane magnetic field direction. This follows 
from the equation~(\ref{equ22}) showing that the current direction is
given by the direction of the in-plane 
nonequilibrium spin which changes upon 
reversing of radiation helicity or the magnetic filed.

\begin{figure}[tb]%%Fig.2
\begin{center}
\centerline{\epsfxsize 75mm\epsfbox{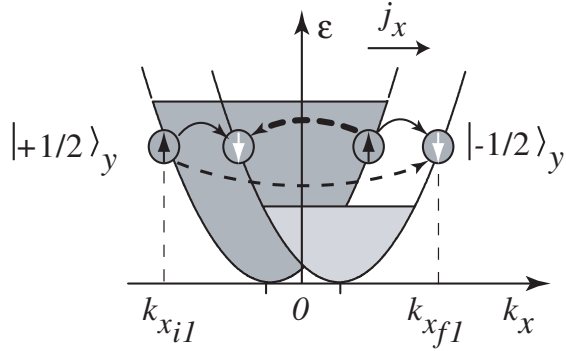}}   %%figure2  figure06_modelsge
\end{center}
\caption{%Microscopic origin of the spin-galvanic current. 
Microscopic origin of the spin-galvanic current in the
presence of $\bm{k}$-linear terms in the electron
Hamiltonian:  %. (angel) One-dimensional sketch: 
the $\sigma_y k_x$ term
in the Hamiltonian splits the conduction band into two parabolas
with the spin $s=\pm 1/2$ pointing in the $y$-direction. If one
spin subband is preferentially occupied, e.g., by spin injection
(the ($|+1/2\rangle_y$-states in the figure) asymmetric spin-flip
scattering results in a current in the $x$-direction. The rate of
spin-flip scattering depends on the value of the initial and final
$\bm{k}$-vectors.  Thus the transitions sketched by broken %dashed
arrows yield an asymmetric occupation of both subbands and hence a
current flow (after Ref.~\protect \cite{Nature02}).
}
\label{spinfig06modelsge}
\end{figure}

Microscopically, the spin-galvanic effect is caused by asymmetric
spin-flip relaxation of  spin polarized  electrons~\cite{Nature02} in systems with
$\bm{k}$-linear contributions to the effective Hamiltonian 
%The band spin splitting $\bm{k}$-linear terms in the effective Hamiltonian 
%$ \hat{H} = \sum_{lm}\beta_{lm}\sigma_l k_m$, where $\bm{k}$ is the electron wave vector, 
${\cal{H}}_k = \sum_{lm}\beta_{lm}\sigma_l k_m$, where $\bm{k}$ is the electron wave vector,
$\sigma_l$ are the Pauli spin matrices and
$\beta_{lm}$ are  real coefficients. The coefficients $\beta_{lm}$
form a pseudo-tensor subjected to the same symmetry restriction as
the transposed pseudo-tensor $\bm{Q}$. The  sources of
Dresselhaus and Rashba $\bm{k}$-linear terms  are the bulk inversion 
 asymmetry~\cite{DK86} and a structural inversion asymmetry~\cite{Bychkov84p78}, respectively. 
% also called the Dresselhaus term~\cite{DK86}
%(including a possible interface inversion asymmetry~\cite{Krebs96p1829}) 
%and a structural inversion asymmetry usually called the Rashba
%term~\cite{Bychkov84p78}.
For a 2DEG system, these terms lead to the situation sketched in
figure~\ref{spinfig06modelsge}. To be specific we show the energy spectrum along $k_x$ with 
the spin dependent term $\beta_{yx}\sigma_y k_x$ of (001)-grown 
zinc-blende structure based  QWs of $C_{2v}$-symmetry and describe 
the corresponding current $j_{x} = Q_{xy}S_y$.

%\begin{figure}[tb]
%\centering
%\includegraphics[width=0.5\linewidth]{figure2.eps}

Spin orientation in $y$-direction causes the
unbalanced population in the  subbands. The current flow is caused
by $\bm{k}$-dependent spin-flip relaxation processes. Spins
oriented in $y$-direction are scattered along $k_x$ from the
higher filled, e.g.,   spin-up subband, $|+1/2 \rangle_y$, to the
less filled  spin-down subband, $|-1/2 \rangle_y$. Four
quantitatively different spin-flip scattering events exist and are
sketched in figure~\ref{spinfig06modelsge}  by bent arrows. The spin-flip
scattering rate depends on the values of the wave vectors of the
initial and the final states~\cite{Averkiev02pR271}. 
%respectively~\cite{Averkiev02pR271}. 
Two scattering processes shown by broken arrows are inequivalent and generate an asymmetric
carrier distribution around the subband minima in both subbands.
This asymmetric population results in a current flow along the
$x$-direction. The uniformity of  spin polarization in space is
preserved during the scattering processes.

\subsection{Zero-bias spin separation and magneto-gyrotropic effects}
\label{IIb}

For linearly polarized radiation magneto-gyrotropic effect due to the spin-galvanic effect 
vanishes because the light absorption does not result in the nonequilibrium spin.
However, in an external magnetic field other kinds of spin photocurrents may be
generated even by unpolarized radiation as it has been proposed
for bulk gyrotropic crystals~\cite{bulli}. 
The microscopic mechanisms of this type spin photocurrents  in 
low-dimensional structures have been developed most recently to describe 
terahertz (THz) radiation induced photocurrents
observed in GaAs-, InAs-, SiGe and GaN-based structures~\cite{Ganichev2006}. 
It has been shown that free carrier absorption of THz radiation results 
in a pure spin current and corresponding spin separation achieved by
spin-dependent scattering of electrons in gyrotropic media. 
The pure spin current in these experiments is converted into an
electric current by application of a magnetic field which
polarizes spins due to the Zeeman effect yielding magneto-gyrotropic effects.

\begin{figure}%% Fig.3
\centerline{\epsfxsize 140mm \epsfbox{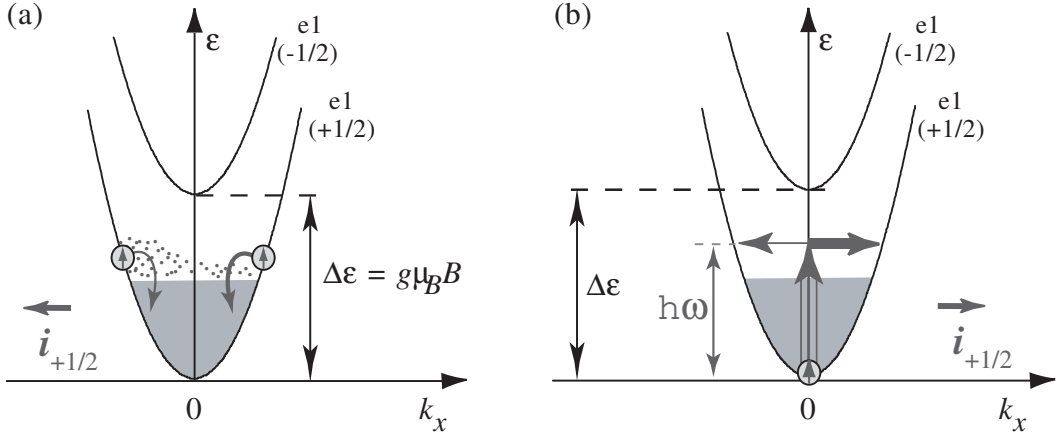}}   %%figure3a_3b  Figure1
\caption{ Microscopic origin of a zero-bias spin separation and
the corresponding magnetic field-induced photocurrent. Zero-bias
spin separation is due to scattering matrix elements linear in
$\bm{k}$ and $\bm{\sigma}$ causing asymmetric  scattering and it
results in spin flows. This process is sketched for the spin-up
subband only and for (a) energy relaxation and (b) excitation via
indirect transitions (Drude-like absorption). Here, scattering is
assumed to have a larger probability for positive $k_x$  than that
for negative $k_x$ as indicated by arrows of different thickness.
Therefore in (a) the energy relaxation rates for positive $k_x$
are larger than  for negative $k_x$ and in (b) the rates of
optical transitions for opposite wave vectors are different. This
imbalance leads to a net spin-up electron flow. In the spin-down
subband the picture is mirror symmetric, resulting in  a net
spin-down electron flow of opposite direction. Thus at zero
magnetic field a pure spin current is generated. The corresponding
electric currents have equal magnitudes and therefore  cancel each
other. An in-plane magnetic field, however, lifts the compensation
of the oppositely directed electron flows yielding a charge
current (after Ref.~\protect \cite{Ganichev06zerobias}). } 
\label{figure1}
\end{figure}

Spin separation due to spin-dependent scattering in gyrotropic
media can be achieved in various ways but all of them must drive
the electron gas into a  nonequilibrium state. One straightforward
method used here is to heat the electron system by %terahertz 
THz
or microwave radiation. Figure~\ref{figure1}(a) sketches the
process of energy relaxation of hot electrons for the spin-up
subband ($s=+1/2$)  in a quantum well containing a two-dimensional
electron gas. Energy relaxation processes are shown by
curved arrows. Usually, energy relaxation via scattering of
electrons is considered to be spin-independent. In gyrotropic
media, like low-dimensional GaAs structures or asymmetric SiGe QWs,
%investigated here, 
however, spin-orbit interaction adds an
asymmetric spin-dependent term to the scattering
probability~\cite{Ganichev06zerobias}. This term in the scattering
matrix element is proportional to components of
$[\bm{\sigma}\times(\bm{k}+\bm{k}^\prime)]$, where $\bm{\sigma}$
is the vector composed of the Pauli matrices, $\bm{k}$ and
$\bm{k}^\prime$ are the initial and scattered electron wave
vectors. %~\cite{footnote1}.
Due to spin-dependent scattering, transitions to positive and
negative $k_x^\prime$-states occur with different probabilities.
Therefore hot electrons with opposite $k_x$ have different
relaxation rates in the two spin subbands. In
figure~\ref{figure1}(a)  this difference is indicated
by arrows of different thickness.
This asymmetry causes an imbalance in the distribution of carriers
 in both subbands ($s=\pm 1/2$) between positive and negative
$k_x$-states.
This in turn yields a net electron flows, $\bm{i}_{\pm 1/2}$,
within each spin subband~\cite{Ganichev06zerobias}. Since the asymmetric part of the
scattering amplitude depends on spin orientation, the
probabilities for scattering to positive or negative
$k^\prime_x$-states are inverted for spin-down and spin-up
subbands.
Thus, the charge currents, $\bm{j}_+ = e\bm{i}_{+1/2}$ and
$\bm{j}_- = e\bm{i}_{-1/2}$, where $e$ is the electron charge,
have opposite directions because $\bm{i}_{+1/2} = -\bm{i}_{-1/2}$
and therefore they cancel each other. Nevertheless, a finite pure spin
current $\bm{J}_{\rm spin} = \frac{1}{2}(\bm{i}_{+1/2} -
\bm{i}_{-1/2})$ is generated since electrons with spin-up and
spin-down move in opposite directions.
This leads to a spatial spin separation and spin accumulation at
the edges of the sample~\cite{footnoteklin}.

As addressed above  pure spin current and
zero-bias spin separation can be converted into a measurable
electric current by application of a magnetic field 
and resulting in the magneto-gyrotropic effects. %~\cite{Ganichev06zerobias}.  
Indeed, in a Zeeman spin-polarized system, the two fluxes $\bm{i}_{\pm 1/2}$,
whose magnitudes depend on the free carrier densities in spin-up
and spin-down subbands, $n_{\pm 1/2}$, respectively, do no longer
compensate each other and hence yield a net electric current (see
figure~\ref{figure1}). For the case, where the fluxes $\bm{i}_{\pm
1/2}$ are proportional to the carrier densities $n_{\pm
1/2}$,
%~\cite{footnote}, 
the charge current is given by
\begin{equation}%% 3
\label{current1}
\bm{j} = e (\bm{i}_{+1/2}  + \bm{i}_{-1/2}) = 4 e S \bm{J}_{\rm
spin} \:,
\end{equation}
where $S = \frac{1}{2}(n_{+1/2} - n_{-1/2})/(n_{+1/2} + n_{-1/2})$
is the magnitude of the average spin. An external magnetic field
$\bm{B}$ results in different populations of the two spin subbands
due to the Zeeman effect. In equilibrium the average spin is given
by
\begin{equation}%%  4
\label{spin}
{\bm{S}} = - \frac{g \mu_B \bm{B}}{4 \bar{\varepsilon}}\:.
\end{equation}
Here $g$ is the electron effective $g$-factor, $\mu_B$  the Bohr
magneton, $\bar{\varepsilon}$  the characteristic electron energy
being equal to the Fermi energy $\varepsilon_F$, or to the thermal
energy $k_B T$, for a degenerate or a non-degenerate 2DEG,
respectively~\cite{Ganichev06zerobias}.

Similarly to the relaxation mechanism, optical excitation of
free carriers by Drude absorption, also involving
electron scattering, is asymmetric and yields spin separation~\cite{Ganichev06zerobias}.
Drude absorption is caused
by indirect intraband optical transitions and includes a momentum
transfer from phonons or impurities  to electrons to satisfy momentum
conservation.
Figure~\ref{figure1}(b)
sketches the process of Drude absorption via virtual states for the
spin-up subband. 
Vertical arrow indicates optical transitions from the initial
state $k = 0$ while the  horizontal
arrows describe an elastic scattering event to a final state with
either positive or negative electron wave vector. 
Due to the spin dependence of scattering, transitions to positive
and negative $k$-states occur with different
probabilities. This is indicated by horizontal arrows of different
thickness. 
The asymmetry causes an imbalance in the distribution of
photoexcited carriers in the subband between
positive and negative $k$-states. 
This in turn yields electron flow. 
Like for relaxation mechanism described above probabilities of scattering 
to positive or negative
$k$ are inverted for spin-down and spin-up subbands,
spin separation takes place and applying
a magnetic field results in a net electric current.

\section{Phenomenological theory}
\label{III}

Phenomenological theory of magneto-gyrotropic effects describes 
dependences of the photocurrent magnitude and direction
on the radiation polarization state and the orientation of the
magnetic field with respect to the crystallographic axes.
Within the linear approximation
in the magnetic field strength $\bm{B}$,  magneto-gyrotropic  %magneto-photogalvanic
effects are given by
\begin{equation} \label{phen0} %% 5
j_\alpha = \sum_{\beta\gamma\delta}
\phi_{\alpha\beta\gamma\delta}\:B_\beta\:\{E_\gamma
E^\star_\delta\} + \sum_{\beta\gamma}
\mu_{\alpha\beta\gamma}\:B_\beta
\hat{e}_\gamma\:E^2_0\:P_{circ}\:.
\end{equation}
Here the fourth rank pseudo-tensor $\bm{\phi}$ is symmetric in the
last two indices, $E_\gamma$ are components of the complex
amplitude of the radiation electric field $\bm E$. In the
following the field is presented as $E = E_0 \bm{e} $ with
$E_0$ being the modulus $|\bm{E}|$ and $\bm{e}$ indicating the
(complex) polarization unit vector, $|\bm{e}| = 1$. The symbol
$\{E_\gamma E^\star_\delta\}$ means the symmetrized product of the
electric field with its complex conjugate,
\begin{equation} \label{sym}
\{E_\gamma  E^\star_\delta\}  = \frac{1}{2}\left(E_\gamma
E^\star_\delta + E_\delta  E^\star_\gamma\right) \:.
\end{equation}
In the second term on the right hand side of equation~(\ref{phen0})
$\bm{\mu}$ is a regular third rank tensor, $P_{circ}$ is the
helicity of the radiation and $\bm{\hat e}$ is the unit vector
pointing in the direction of light propagation. While the second
term in equation~(\ref{phen0}) requires circularly polarized radiation 
and represents the spin-galvanic effect, the first term may be
non-zero even for unpolarized radiation.

For (001)-oriented asymmetric QWs based on zinc-blende lattice
III-V %or II-VI 
compounds and belonging to $C_{2v}$ point group
the phenomenological
equation~(\ref{phen0}) for  the magneto-photogalvanic
effects induced by normally-incident radiation reduces
to~\cite{Belkov05p3405}
\begin{eqnarray} %% 7
\label{phen} 
j_{x} = S_1  B_{y}I + 
S_2 B_{y} \left( |e_{x}|^2 - |e_{y}|^2 \right) I + %\nonumber\\
S_3 B_{x} \left( e_{x} e^*_{y} + e_{y} e^*_{x} \right) I  +  
S_4 B_{x}   I P_{circ}\:,      \nonumber\\
j_{y} = S'_1  B_{x}I + 
S'_2 B_{x} \left( |e_{x}|^2 - |e_{y}|^2 \right) I + %\nonumber\\
S'_3 B_{y} \left( e_{x} e^*_{y} + e_{y} e^*_{x} \right) I  +  
S'_4 B_{y}   I P_{circ}\:,
\end{eqnarray}
where, for simplicity, we set for the intensity $I=E_0^2$. Here the
parameters $S_1$ to $S'_4$ correspond to the non-zero components 
of the tensors $\bm{\phi}$ and $\bm{\mu}$ allowed by the $C_{2v}$ 
point group and only in-plane
components of the magnetic field are taken into account. 
%For the current in $y$ direction $x$ and $y$ should be interchange and 
%parameters $S_1$ to $S_4$ replaced by parameters $S'_1$ to $S'_4$. 

For linearly polarized radiation at normal incidence and ${\bm{B}} \parallel y$ we have 
\begin{eqnarray} %% 8 and 9
\label{NatPhys}
j_{x} = S_1 B_{y} I + S_2 B_{y} I \cos{2 \alpha} \:,\: \nonumber\\
j_y = S'_3 B_{y} I \sin{2 \alpha}\:,
\end{eqnarray}
where  $\alpha$ is angle between the  radiation polarization 
vector $\bm{e}$ and $y$-axis.

If in experiments an elliptically polarized radiation is used the
spin-galvanic effect described by the last terms in equations~(\ref{phen}) 
can also be non-zero. A convenient way of variation of polarization state is
passing laser radiation, initially linearly polarized along ${x}$-axis, 
through a $\lambda/4$-plate. Rotation of the plate by an 
angle $\varphi$  changes helicity and the azimuth angle of the ellipse. The helicity 
$P_{circ}$ of the incident light varies from $-1$ (left
handed, $\sigma_-$) to $+1$ (right handed, $\sigma_+$) according
to $P_{circ} = \sin{2 \varphi}$ and the degree of linear polarization $P_{\rm lin} = \sin{4 \varphi}/4$.
%Check self-consistence in the whole text 2 or 4!!!
The total current for  normal incidence  and ${\bm B} \parallel y$ is described in this case by
\begin{eqnarray}
\label{NatPhys2}
j_{x} = S_1 B_{y} I + S_2 B_{y} I (1 +  \cos4 \varphi)/2  \:,\: \nonumber\\
j_y = S'_3 B_{y} I \sin{4 \varphi}/2  + S'_4 B_{y} I \sin{2 \varphi} \:.
\end{eqnarray}
It is seen that all four contributions are characterized by
different dependences of the photocurrent magnitude and direction
on the radiation polarization state and the orientation of the
magnetic field with respect to the crystallographic axes. As a
consequence, a proper choice of experimental geometry allows one to
investigate each partial photocurrent separately.

\section{Methods }
\label{IV}

%\subsection{Experimental technique}
%\label{IVB}

For  optical excitation mid-infrared, terahertz and
visible laser radiation was used. Most of the measurements were
carried out in the infrared with photon energies less than the
energy gap of  investigated semiconductors. For investigations of
spin photocurrents infrared excitation has several advantages.
First of all below the energy gap the absorption is very weak and
therefore allows homogeneous excitation with marginal heating of
the 2DEG. Furthermore, in contrast to inter-band
excitation, there are no spurious photocurents due to other
mechanisms like the Dember effect, photovoltaic effects at
contacts and Schottky barriers etc. Depending on the photon energy
and QW band structure the mid-infrared and THz radiation induce direct
optical transitions between size quantized subbands 
or, at longer wavelength, indirect  optical
transitions (Drude absorption) in the lowest subband.

A high power pulsed mid-infrared  transversely excited
atmospheric pressure carbondioxide (TEA-CO$_2$) laser and a
molecular terahertz laser~\cite{Ganichev2006}
have been used  as radiation sources in the spectral range between
9.2~$\mu$m and 496~$\mu$m with power  $P \simeq $ 5~kW. 
The corresponding photon energies $\hbar \omega$  lie in the range of 135~meV to 2~meV. 

Optically pumped  molecular lasers emit linearly polarized radiation 
those orientation
is determined by the polarization of the pump radiation.
In experiment the plane of polarization on the sample
is rotated applying  $\lambda/2$-plates which enable one to vary the azimuth
angle $\alpha$ between $0^\circ$ and $180^\circ$ corresponding to all possible orientations of 
the electric field vector in the ($xy$) plane. Here $\alpha = 0$
is chosen so that the  radiation polarization vector on the sample is directed along $y$-axis.

To investigate magneto-gyrotropic effects due to the spin-galvanic effect 
an elliptically, in particularly circularly, polarized light is required.
The  polarization of the laser beam in this case is modified from linear 
to elliptical applying crystal quartz $\lambda/4$-plates. 
Usually $\varphi = 0$ is chosen for optical axis of 
the quarter-wave plate coinciding with the incoming laser polarization vector. 
%By that the linear polarization degree is given by $P_{lin} = \sin{4 \varphi}/2$.

\begin{figure}
\centerline{\epsfxsize 86mm \epsfbox{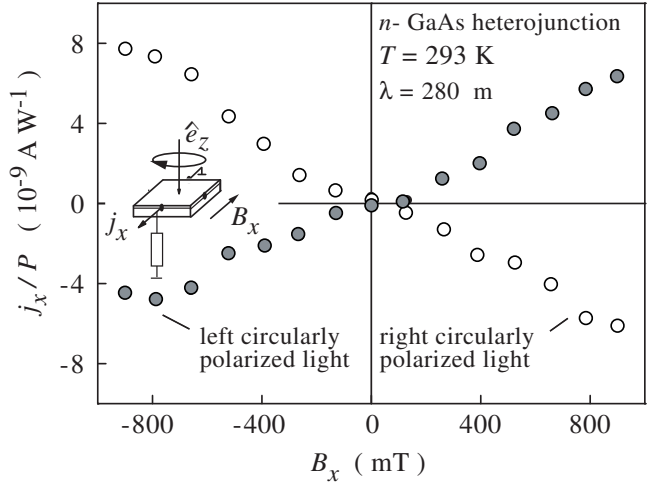}}  %%figure4  spinfig17sgemagfieldRT
\caption{Magnetic field dependence of the spin-galvanic current normalized by
$P$ achieved by intra-subband transitions within $e${\it 1} conduction
subband by excitation with  radiation of $\lambda~=$~280~$\mu$m
wavelength. Results are plotted for an (001)-grown GaAs/AlGaAs single
heterojunction at room temperature (after Ref.~\protect\cite{Ganichev03p935}).}
\label{spinfig17sgemagfieldRT}
\end{figure}

Magneto-gyrotropic photogalvanic effects have been investigated 
on a large variety of low-dimensional structures comprising 
GaAs-, InAs-, SiGe- and GaN-based heterostructures. As it follows from the equation~(\ref{phen0})
the relative direction between the current and the magnetic field and 
polarization dependences of various current contribution can be different for various 
symmetry point groups. Mostly frequently investigations have been  carried out on 
structures describing by two point groups: $C_{2v}$ and $C_{s}$. 
Higher symmetric structures ($C_{2v}$) were (001)-oriented asymmetric QWs and 
(110)-grown symmetric GaAs QWs.  
Structures belonging to lower symmetry class $C_s$ were (110)-oriented asymmetric GaAs QWs. 
Besides these structures magneto-gyrotropic effects were observed in (001)-grown SiGe QWs of $C_{2v}$ point group and 
(0001)-grown GaN/AlGaN heterojunctions belonging to $C_{3v}$-symmetry.
MGE has been investigated at room temperature or mounted in an optical
cryostat which allowed the variation of temperature in the  range
of 4.2~K to 293~K. The  photocurrent $j$ was measured in the
unbiased structures via the voltage drop across a 50~$\Omega$ load
resistor in a closed circuit configuration %~\cite{APL00} 
(see
inset in figure~\ref{spinfig17sgemagfieldRT}). An external magnetic field for room temperature
is obtained by a conventional electromagnet with the magnetic field up to
1~T and at 4.2~K using a superconducting split-coil magnet with $B$ up to 3~T.

\section{Experimental results and discussion}
\label{V}

\subsection{Magneto-gyrotropic effect due to the Larmor precession induced spin-galvanic photocurrent}
\label{SGEoptmag}

Magneto-gyrotropic effect due to the Larmor precession induced spin-galvanic photocurrent
has been observed for (001)-grown $n$-type GaAs and InAs QWs as well as for (0001)-grown GaN/AlGaN structures 
applying both, visible and THz radiation~\cite{Nature02,PRB03sge}.
This effect has been described in details in several recent reviews~\cite{Ganichev2006,Ivchenkobook2,Ganichev03p935} and 
we will address here only main features of this phenomena and discuss its application to
investigation of inversion asymmetry in quantum well structures.

\begin{figure}  %%  5
\centerline{\epsfxsize 86mm \epsfbox{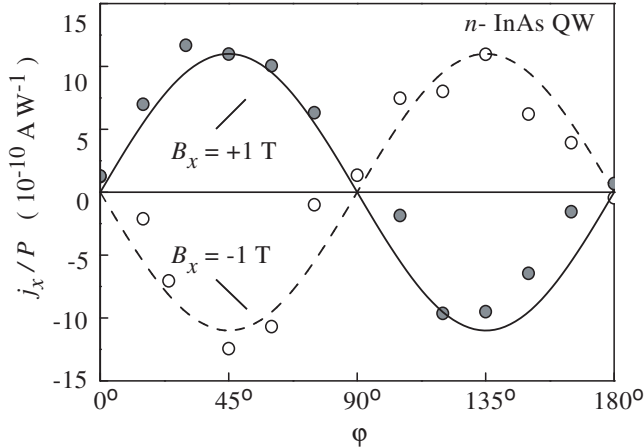}}   %%figure5  spinfig19sgephi
\caption{Spin-galvanic current   normalized by $P$ as a function
of the phase angle $\varphi$ in an (001)-grown $n$-type InAs QW of
15~nm width at $T$~=~4.2~K. The photocurrent  excited by normal
incident radiation of $\lambda~=~148$~$\mu$m is measured in
$x$-direction parallel (full circles) and anti-parallel (open
circles) to the in-plane magnetic field $B_x$. 
Solid and dashed curves are fitted 
to $j_x \propto \pm \sin{2 \varphi}$ (after Ref.~\protect \cite{PASPS02sge}).
%after equations~(\protect \ref{equ32})  using the  %%%  !!!!!
%same scaling of the ordinate. 
}
\label{spinfig19sgephi}
\end{figure}

Most of experiment has been carried out applying experimental geometry 
described in section~\ref{IV} and sketched in  figure~\ref{spinfig07larmor}.
The magneto-gyrotropic effect due to spin-galvanic effect manifest itself by 
the presence of the current contribution $j \propto B P_{circ}$,
which reverses its direction upon both, switching the radiation handedness for 
a fixed magnetic field, and 
changing of the in-plane magnetic field direction at fixed radiation 
helicity. Typical helicity dependence for two directions of magnetic fields 
is shown in figure~\ref{spinfig19sgephi}. 
For low magnetic fields $B$, where $\omega_L\tau_s < 1$ holds, the
current  increases linearly as expected from equations~(\ref{equ22})
and~(\ref{Hanle}).  This is seen in the room
temperature data of figure~\ref{spinfig17sgemagfieldRT}. The
polarity of the current depends on the direction of  the excited
spins ($S$ aligned along $\pm z$-direction for right or 
left  circularly polarized light, respectively)
and on the direction of the applied magnetic field ($\pm B_x$-direction)
which determines the Larmor precession direction 
(see figures~\ref{spinfig17sgemagfieldRT} and~\ref{spinfig19sgephi}). 
For magnetic field applied along $\langle 110\rangle$, as shown in these figures, 
the spin-galvanic current is parallel (anti-parallel) to the magnetic field vector.  
For $\bm{B} \parallel \langle 100\rangle$ both the transverse and the
longitudinal effects are observed~\cite{Ganichev03p935}.

In some experiments, carried out at low temperature 
and for higher magnetic fields, it is observed that 
with rising magnetic field strength 
the current assumes a maximum and decreases upon further increase of $B$~\cite{Nature02}. 
This drop of the current is ascribed to the
Hanle effect~\cite{Meier}. The experimental data are well
described by %Eqs.~(\ref{equ23}) and (\ref{equ30}). The observation
equation~(\ref{Hanle}). The observation
of the Hanle effect demonstrates that free carrier intra-subband
transitions can polarize the spins of electron systems. The
measurements  allow one to obtain the spin relaxation time $\tau_s$
from the peak position of the photocurrent where $\omega_L\tau_s
=1$ holds~\cite{Nature02}.

We note that the observation of the mid-infrared and terahertz radiation excited spin-galvanic
effect, which is due to spin orientation, gives clear evidence
that direct intersubband and Drude absorption of circularly
polarized radiation result in a monopolar spin
orientation. Mechanisms of the monopolar spin orientation were analyzed 
in~\cite{PASPS02monop,Ivchenko04p379}. 
 
The microscopic theory of the SGE in QWs was developed in~\cite{PRB03sge,GolubSGE}.
Within the model of elastic scattering the
current is not spin polarized since the same number of spin-up and
spin-down electrons move in the same direction with the same
velocity. The spin-galvanic current 
%
%Intersubband or generally? Check it!
%
can be estimated by~\cite{PRB03sge}
\begin{eqnarray}\label{kinetic}  %%%  12
j_x
& = & Q_{xy}S_y \sim e\: n_s
\frac{\beta_{yx}}{\hbar} \frac{\tau_p }{\tau^\prime_s}
S_y\:,
\end{eqnarray}
and the similar equation for $j_y$, where $n_s$ is the 2DEG
density, $\tau^\prime_s$ is the spin relaxation time due to the
Elliott--Yafet mechanism~\cite{Meier}.   
Since spin-flip scattering is the origin
of the current given by equation~(\ref{kinetic}), this
equation is valid even if D'yakonov--Perel'
mechanism~\cite{Meier,DP71}   
of spin relaxation dominates. The
Elliott--Yafet relaxation time $\tau^\prime_s$ is proportional to
the momentum relaxation time $\tau_p$. Therefore the ratio $\tau_p
/ \tau_s^\prime$ in equation~(\ref{kinetic}) does not
depend on the momentum relaxation time. The in-plane average spin,
e.g., $S_y$ in equation~(\ref{kinetic}), decays with the total spin
relaxation time $\tau_s$ and, hence, the time decay of the
spin-galvanic current following pulsed photoexcitation is
described by the exponential function $\exp{(- t/ \tau_s)}$. In
contrast, the circular PGE current induced by a short-pulse decays
within the momentum relaxation time $\tau_p$ allowing to
distinguish these two effects in time resolved measurements.
In general, in addition to the kinetic contribution to the current
there exists the so-called relaxational contribution which arises
due to the ${\bm{k}}$-linear terms neglecting the Elliott--Yafet
spin relaxation, i.e., with allowance for the D'yakonov--Perel'
mechanism only~\cite{Ivchenkoprivatcommunication}.

\begin{figure}[t]  %%%Fig.6
\centering
\centerline{\epsfxsize 140mm \epsfbox{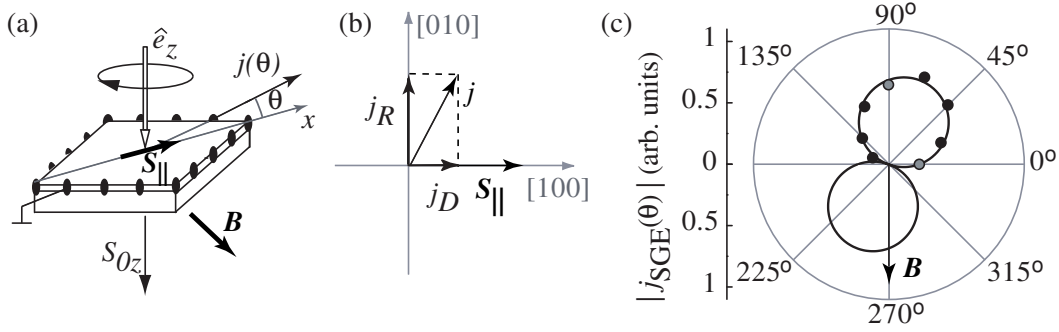}}%%figure6a_6c  figure11_RD
%\picplace{5cm}{2cm} % Give the correct figure height and width in cm
\caption[]{The separation of the Dresselhaus and Rashba
contributions to the spin-galvanic effect observed in an $n$-type InAs single QW at room
temperature for the case of the electron spin ${\bm S}_{\parallel} \parallel [100]$.
(a) Geometry of the experiment. (b) The direction of Dresselhaus and Rashba contributions to
the photocurrent. (c) The spin-galvanic current measured as a function of the angle $\vartheta$
between the pair of contacts and the $x$-axis (after Ref.~\protect\cite{PRL04}). }
\label{figure11_RD}       % Give a unique label
\end{figure}

An important application of the spin-galvanic is addressed
in~\cite{PRL04}. It is  demonstrated that angular dependent
measurements of spin photocurrents allow one to separate the
Dresselhaus and Rashba terms.
These experiments were carried out on (001)-oriented QWs for which %the
linear in wave vector part of Hamiltonian %equation~(\ref{lkterms}) 
for the first subband reduces to
\begin{eqnarray} \label{biasia}  %%  13
%\fl 
{\cal{H}}^{(1)}_{\bm{k}} &=& \alpha(\sigma_{x_0} k_{y_0} - \sigma_{y_0} k_{x_0}) +
\beta(\sigma_{x_0} k_{x_0} - \sigma_{y_0} k_{y_0})
 \:,
\end{eqnarray}
where the parameters $\alpha$ and $\beta$
result from the structure-inversion and bulk-inversion
asymmetries, respectively, and $x_0, y_0$ are the crystallographic
axes $[100]$ and $[010]$. 
%
%The next sentence is already introduced above somehow may we should delete it
Note that, in the coordinate system with
$x\parallel [1\bar{1}0]$ %,\quad 
 and $y\parallel [110]$, the matrix $
{\cal{H}}^{(1)}_{\bm k}$ gets the form $\beta_{xy} \sigma_x k_y +
\beta_{yx}  \sigma_y k_x$ with $\beta_{xy} = \beta + \alpha$,
$\beta_{yx} = \beta - \alpha$. 
According to equation~(\ref{kinetic}) the current components $j_x$, $j_y$ are proportional,
respectively, to $\beta_{xy}$ and $\beta_{yx}$ 
%Here should be x_0 and y_0, please check!
and, therefore, angular dependent measurements of spin
photocurrents allow one to separate the Dresselhaus and Rashba
terms. By mapping the magnitude of the spin photocurrent in the QW
plane the ratio of both terms can directly be determined from
experiment and does not rely on theoretically obtained quantities.
The relation between the photocurrent and spin directions can be
conveniently expressed in the following matrix form
\begin{equation}  %% 14
\label{bissiam} \mbox{\boldmath$j$}  \propto \left(
{{\begin{array}{*{20}c}
 \beta \hfill  & \,\,\, -\alpha \hfill \\
  \alpha  \hfill  & \,\,\, - \beta \hfill \\
\end{array} }} \right)\mbox{\boldmath$S_\parallel$} \:,
\end{equation}
where $\mbox{\boldmath$j$}$ and $\mbox{\boldmath$S_\parallel$}$ 
are two-component columns with the
in-plane components along the crystallographic axes $x_0 \parallel
[100]$ and $y_0 \parallel [010]$. The directions of the
Dresselhaus and Rashba coupling induced photocurrents are shown in
figure~\ref{figure11_RD}(b) for the particular case ${\bm S}_{\parallel} \parallel [100]$.

% For figures use

Figure~\ref{figure11_RD}(c) shows the angular dependence of the
spin-galvanic current $j(\vartheta)$
measured on $n$-type (001)-grown InAs/Al$_{0.3}$Ga$_{0.7}$Sb
single QW  of 15\,nm width at room temperature. Because of the
admixture of photon handedness-independent magneto-gyrotropic
effects (see subsection~\ref{Vb}) the spin-galvanic effect is
extracted after eliminating this current contributions:  
%which are helicity-independent: 
$j =\left(j_{\sigma_+}-j_{\sigma_-}\right)/2$.

The nonequilibrium in-plane spin polarization {\boldmath$S_\parallel$} is prepared as
described in section~\ref{IV} (see also
figure~\ref{figure11_RD}(a)). The angle between the magnetic field and
{\boldmath$S_\parallel$} can in general depend on details of the
spin relaxation process. In these particular InAs QW structures
the isotropic Elliott--Yafet spin relaxation
mechanism dominates. Thus, the in-plane spin polarization
{\boldmath$S_\parallel$} is always perpendicular to {\boldmath$B$}
and can be varied by rotating {\boldmath$B$} around $z$ as
illustrated in figure~\ref{figure11_RD}(a). The circle in
figure~\ref{figure11_RD}(c) represents the angular dependence
$\cos{(\vartheta - \vartheta_{\rm max})}$, where
$\vartheta$ is the angle between the pair of contacts
and the $x$-axis and $\vartheta_{\rm max} = \arctan{j_R/j_D}$.
The best fit in this sample is achieved for the ratio
$j_R/j_D=\alpha/\beta = 2.1$. The method was also used for
investigation of Rashba/Dresselhaus spin splitting in GaAs/AlGaAs
heterostructures~\cite{PRB07gig} where spin relaxation  is controlled by
D'yakonov--Perel' mechanism. These experiments
demonstrate that growth of structures with various $\delta$-doping
layer position accompanied by experiments on spin-galvanic effect
makes possible a controllable variation of the structural inversion
asymmetry and preparation of
samples with equal Rashba and Dresselhaus constants or with
a zero Rashba constant.

The effect inverse to the spin-galvanic effect is the electron 
spin polarization generated
by a charge current ${\bm j}$. It was predicted in
\cite{Ivchenko78p640} and observed in bulk tellurium
\cite{Vorobjev79p441}.
%Aronov, Lyanda-Geller~\cite{Aronov89p431},
%Edelstein~\cite{Edelstein89p233} and Vas'ko, Prima~\cite{Vasko}
It was futher demonstrated that spin orientation by current 
is also possible in QW systems~\cite{Aronov89p431,Edelstein89p233,Vasko}.
This study was extended in
Refs.~\cite{Aronov91p537,Chaplik,VaskoRai,TarJETP,Schlieman07,Raichev}.
Most recently the first direct experimental proofs of this effect
were obtained in semiconductor
QWs~\cite{Ganichev04p0403641,Silov2004}
as well as in strained bulk material~\cite{Kato04p176601}.
At present inverse spin-galvanic effect has been observed in
various low-dimensional structures based on GaAs, InAs, ZnSe and
GaN applying various experimental techniques, comprising transmission of
polarized THz radiation, polarized photoluminescence and space resolved
Faraday rotation~\cite{Ganichev04p0403641,Silov2004,Kato04p176601,Silov04,Sih05isge,Stern06isge,Yang06}.

\begin{figure}[width=4cm] %% 7
\centerline{\epsfxsize 70mm \epsfbox{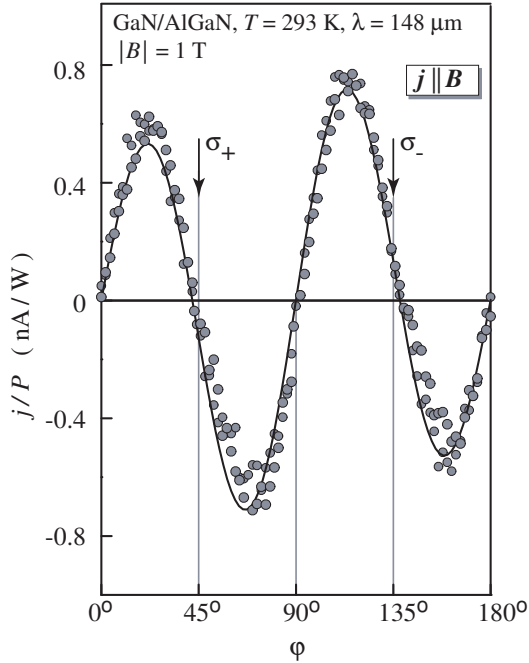}} %%figure7  figure3beatingGaN
\caption{Magnetic field induced photocurrent $J$ as a function of the phase angle $\varphi$
defining the radiation helicity.
The photocurrent signal is measured in GaN/AlGaN heterojunction at room temperature in the longitudinal
geometry $\bm j \parallel \bm B$     %% $\bm j \parallel \bm B$   ???????
under normal incidence of the radiation with $P \approx 10$~kW. The full line is the
fit after the second equation of equations~\protect(\ref{NatPhys2}). After Ref.~\protect\cite{Weber08}.
}
\label{figure3beatingGaN}
\end{figure}

In some structures, however, the helicity dependence of the photocurrent 
is  not as simple as shown in figure~\ref{spinfig19sgephi}, where current 
is proportional to the radiation helicity. It has been observed that
the dependence of the magneto-induced photocurrent 
can show beatings caused by interplay of two terms, one being
$\propto\sin 2\varphi$ and the other $\propto \sin 4 \varphi$. The latter term stems for the MGE 
due to zero-bias spin separation and in some structures can even play a dominant role 
%completely hiding 
overweighting the spin-galvanic effect. Such an interplay is shown in figure~\ref{figure3beatingGaN} 
for GaN/AlGaN heterojunction for photocurrent measured along magnetic field. 
Similar dependences  have been observed in GaAs- and InAs-based  QWs. %samples.
The contribution $\propto \sin 4 \varphi$ directly follows form the phenomenological theory (see equation~(\ref{phen0}))
and corresponds to the degree of linear polarization describing by the Stokes parameter $P_{lin} = \sin 4\varphi / 4$.
%?? /4
It can be investigated without admixture of the spin-galvanic effect applying linearly polarized radiation. 
In the next subsection we consider details of this  type of magneto-gyrotropic effects.

\subsection{Magneto-gyrotropic effects due to the zero-bias spin separation}
\label{Vb}

\begin{figure}  %% 8
\centerline{\epsfxsize 100mm \epsfbox{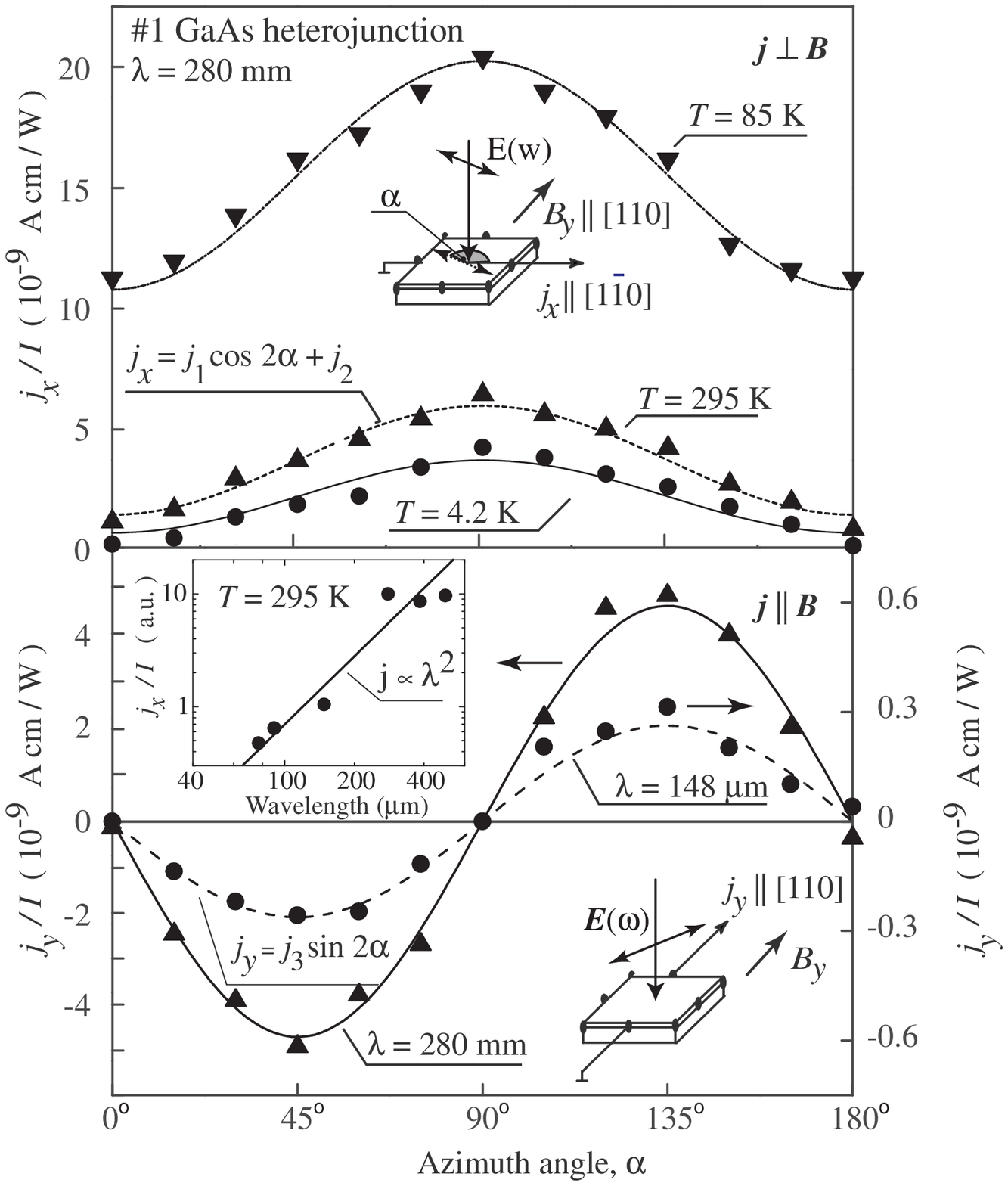}}  %figure8  Figure3_alpha
\caption{
 Photocurrents for GaAs/AlGaAs heterojuncton %(sample~1)
as a function of angle $\alpha$. Data are obtained at
$B_y$~=~0.3~T. Upper panel: photocurrent $\bm j \, \bot \, \bm B
\,\|\,y$ at  $\lambda = 280$~$\mu$m and T~=~4.2~K, 85~K and 295~K.
Lines are fits according to $j_x= j_1 \cos 2\alpha + j_2$.  
%(see first equation of equations~\protect(\ref{NatPhys})).
%, ~\protect(\ref{polariz1}) and~\protect(\ref{trigon}). 
Lower panel: photocurrent $\bm j \, \|
\, \bm B \,\|\,y$ measured at room temperature for $\lambda =$~148
and 280~$\mu$m. Lines are fitted to $j_y = j_3 \sin 2\alpha$.
%(see second equation of equations~\protect(\ref{NatPhys}))
%, see equations~\protect(\ref{polariz2}) and~\protect(\ref{trigon}). 
Insets
show the experimental geometries. An additional inset in the lower
panel displays the wavelength dependence of the signal for
transverse geometry, the full line shows $j_x \propto\lambda^2$. From Ref.~\protect\cite{Ganichev06zerobias}.}
 \label{Figure3_alpha}
\end{figure}

Magneto-gyrotropic effects  due to the zero-bias spin separation 
do not require circularly polarized light %radiation 
and can be observed applying linearly polarized or even unpolarized radiation. 
As in the previous case of the spin-galvanic effect the current pulse follows the excitation pulse shape and 
its direction reverses upon the change of the magnetic field direction,
however it does not depend on the radiation handedness.
Figure~\ref{Figure3_alpha} shows a typical variation the photocurrent upon 
changes of the azimuth angle $\alpha$ for the in-plane magnetic field aligned along $y$-axis, 
$B_y$. The data are obtained for (001)-grown  GaAs/AlGaAs heterojuncton %GaAs QW 
at normal incidence with terahertz radiation resulting in the 
Drude-like absorption~\cite{Ganichev06zerobias}. The polarization dependence of the current $\bm{j}$ 
is good described by the phenomenological equations~(\ref{NatPhys})
and correspondingly can be well fitted in  the transverse geometry  
by $j_x = j_1 \cos 2 \alpha + j_2$, and by $j_y = j_3 \sin 2 \alpha $  for the
longitudinal geometry. In the following we will use following definitions $j_1 = S_2 B_{y} I$,
$j_2 = S_1 B_{y} I$,   $j_3 = S'_3 B_{y} I$.
We note that for the photocurrent detected in the direction perpendicular to the 
magnetic field a polarization independent 
offset is observed, demonstrating that this effect can appear at the excitation with 
unpolarized radiation.

The variation of the photocurrent strength and direction upon changing of the 
azimuth angle is characteristic for MGE due to Drude absorption~\cite{Belkov05p3405}.
In this case the polarization dependences of the photocurrent remain  the same, 
independent of temperature, wavelength and  material.
An increased wavelength at constant intensity results in an increased signal strength only. The
wavelength dependence for both configurations
is described by $j \propto \lambda^2$ for the
wavelengths used (see inset in
figure~\ref{Figure3_alpha}, lower panel) and reflects the spectral behaviour of  Drude
absorption, $\eta(\omega) \propto 1/\omega^2$ at $\omega \tau _p \gg 1$ (see~\cite{Seeger}).
Here $\eta(\omega)$ is the 2DEG's absorbance at frequency $\omega$.
For  intersubband transitions excited by the mid-infrared radiation MGE has also been observed. 
In this case the photocurrent is detected in the direction perpendicular to the magnetic field 
only and does not show any polarization dependence. This result is not surprising because, 
in contrast to Drude mechanism, absorption due to direct
intersubband transitions does not rely on scattering and the photoexcitation mechanism 
of the photocurrent caused is absent.

\begin{figure}  %%   9
\centerline{\epsfxsize 80mm \epsfbox{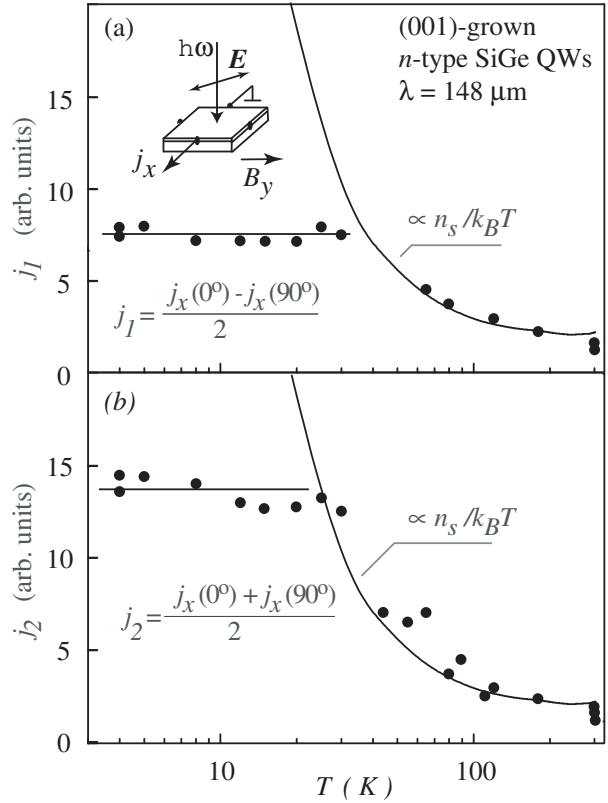}}  %%figure9a_9b  fig9_13_temper
\caption{
 Temperature dependences of the
contributions $j_1$ (a) and $j_2$ (b) to the photocurrent $j_x$ in a
magnetic field ${\bm B} \parallel y$ ($B = 0.6$~T).
Full lines are fits to $A n_s /k_BT$
with a single fitting parameter $A$,
and to a constant.  %, respectively.
>From Ref.~\protect\cite{SiGe07}.}
 \label{figure16_MGEcombined}
\end{figure}

The key experiment supporting microscopic mechanisms discussed in subsection~\ref{IIb} 
is investigation of the temperature dependence of the photocurrent.
The analysis shows~\cite{SiGe07} that for fixed polarization for both excitation,
and relaxation mechanisms (figures~\ref{figure1}(a) and (b)) the 
current is proportional to the frequency dependent absorbance $\eta(\omega)$,
momentum relaxation time $\tau_p$, light intensity $I$ and
average spin $S$: $j\propto \eta(\omega) I \tau_p S$.
Such type of expression for the temperature dependence 
is valid for fixed scattering mechanism, e.g., phonon or impurity scattering.
For Drude absorption  the temperature dependence of the current can be reduced to $n_s S$.
Since, Drude absorption, $\eta (\omega ) \propto n_s /\tau _p $ at $\omega \tau _p \gg 1$ (see~\cite{Seeger})
and at low temperatures $S \propto 1/\varepsilon _F \propto
1/n_s $ (see equation~(\ref{spin})), the current $j / I \propto \tau _p \,\eta (\omega )
S$ is constant and independent of $\tau _{p}$ and $n_{s}$. 
At high temperatures $S$ is sufficiently well described by the Bolzmann distribution and hence $S \propto 1/k_B T$,
see equation~(\ref{spin}). Therefore, the current $j$ is proportional to $n_s /T$
and becomes temperature dependent, concordant with experiment.
An example of such a temperature behaviour is shown in 
figure~\ref{figure16_MGEcombined}  where the
temperature dependences of the currents $j_1$ and $j_2$
corresponding to the excitation and relaxation mechanisms is depicted.
%in figures~\ref{figure16_MGEcombined}(a) and \ref{figure16_MGEcombined}(b), respectively. 
The data are obtained in an $n$-type SiGe QW structure for the magnetic field 0.6~T
under excitation with %a THz molecular laser ($\lambda = 148$~$\mu$m). 
radiation of $\lambda = 148$~$\mu$m wavelength.

\begin{figure}  %% 10
\centerline{\epsfxsize 100mm \epsfbox{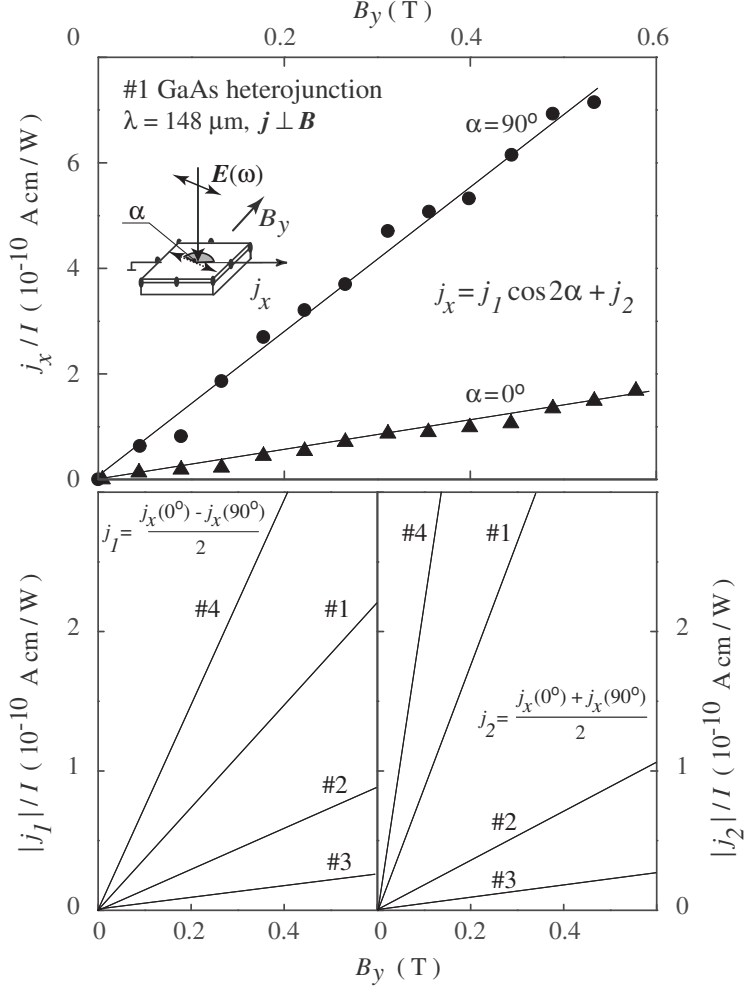}}  %%figure10  Figure4_asymmetry
\caption{
Magnetic field dependences of the transverse photocurrent.
Upper panel: $j_x(B)$ normalized by radiation intensity $I$,
measured for GaAs-based heterojunction (sample~1) at room temperature for $\alpha = 90^\circ$
and $\alpha = 0^\circ$. Lower panels: $j_1(B)$ (left panel) and
$j_2(B)$ (right panel) obtained by subtracting and  adding the
currents for the two polarization directions for samples~1-4 (after  Ref.~\protect\cite{Ganichev06zerobias}).  }
 \label{Figure4_asymmetry}
\end{figure}

The theory of the magneto-gyrotropic effect due to zero-bias spin separation caused 
by Drude absorption and relaxation of heated electron gas has been developed in~\cite{Ganichev06zerobias}.
Besides characteristic temperature and polarization dependences this theory demonstrates that the 
photocurrent is proportional to the degree of inversion asymmetry. This is a natural results because 
the asymmetry of scattering is due to BIA/SIA. The experiments carried out on  samples with different degree of asymmetry
shows that the current strength is indeed proportional to the degree of structural asymmetry. Moreover photocurrent 
reverses its direction upon reversing of sign of the SIA contribution.
This is demonstrated by the figure~\ref{Figure4_asymmetry} which shows $j_1$ and $j_2$ 
attributed to the photoexcitation (figure~\ref{figure1}(b)) and to the relaxation (figure~\ref{figure1}(a)) mechanisms, respectively.
Due to the larger $g$-factor of sample~4 (InAs QW),
causing larger average spin  $S$, the currents are
 largest for this structure.
The other three samples are GaAs-based heterostructures which differ
in structural inversion asymmetry. Sample~1 is a heterojunction %(see Table~I)
which, due to the triangular confinement potential, is expected to have the strongest SIA contribution.
Samples~2 and~3 are quantum wells of the same width, asymmetrically and symmetrically modulation
doped, with larger and smaller strength of SIA,
respectively.
The fact that with decreasing strength of the SIA coupling coefficient
(from sample~1 to~3) the
currents become smaller is in excellent agreement with our picture
of asymmetric scattering  driven currents.
%The coupling strength constant controls
%the current via $V_{\alpha \beta}$ in equations~(\ref{polariz1}), (\ref{polariz2}) and
%equivalent expressions for other scattering mechanisms: The larger  the coupling strength
%the larger is the effect of asymmetric scattering.   ???????????????

\subsection{Magneto-gyrotropic effect in (110)-grown quantum wells}

Most recently magneto-gyrotropic effects have been applied to conclude on 
symmetry and spin dephasing in (110)-grown quantum wells.
Quantum wells  on (110)-oriented GaAs substrates attract 
growing attention in spintronics due to their extraordinary 
slow spin dephasing~\cite{Ohno1999,Harley03,x3,x4}. 
The reason for the long spin lifetime of several nanoseconds is the (110) crystal orientation: 
Then the effective magnetic field due to spin-orbit coupling points into 
the growth direction~\cite{DK86} and spins oriented along this direction do not precess. 
Hence the D'yakonov--Perel' spin relaxation mechanism~\cite{DP71} which is based 
on the spin precession in the effective magnetic field and
usually limits the spin lifetime of conduction electrons is suppressed.
If, however, QWs are asymmetric, the structural inversion symmetry is broken and 
Rashba spin-orbit coupling causes an in-plane effective magnetic field, 
thus speeding-up spin dephasing. 
To judge the symmetry of QWs, one has to rely on 
the  growth process but no independent method 
to check the structure symmetry is readily available. 
The magneto-gyrotropic effect %(MGE)~\cite{Belkov05p3405} 
is an ideal tool to probe the  symmetry of (110)-grown QWs. 
The photocurrent is only observed for asymmetric structures but vanishes 
if QWs are symmetric. This statement has also been supported by 
time-resolved Kerr rotation  that the spin relaxation time is maximal 
in QW which does not show MGE.

\begin{figure}%[t]%[width=4cm]   %% 11
\centerline{\epsfxsize 86mm \epsfbox{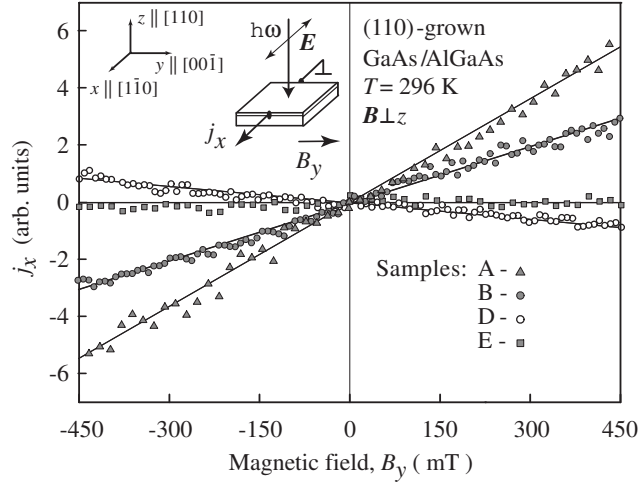}}  %   figure11  fig2_inplane
\caption{Magnetic field dependences of $j_x$ 
for the radiation polarized along $x$-axis and 
an in-plane magnetic field, $\bm{B} \parallel y$ (after  Ref.~\protect\cite{condmat110}).}
\label{fig2_inplane}
\end{figure}

In~\cite{condmat110} the structural inversion asymmetry has been varied by 
the  $\delta$-doping position in respect to the QW. 
Quantum wells differ essentially  in their doping profile:
Sample A is a single heterojunction and 
has the strongest asymmetry 
stemming from the triangular confinement potential. 
In samples B and D, the doping layers are
asymmetrically shifted off the barrier center either to the left  or to
the right, respectively.  This asymmetric doping
yields an asymmetric potential profile inside the QWs. 
Samples grown along $z\parallel[110]$ were square shaped 
with the sample edges  of 5~mm length oriented along 
$x\parallel[1{\bar 1}0]$ and $y\parallel[00{\bar 1}]$.
The degree of  SIA is reflected in the magnetic field 
dependence of the photocurrent displayed in figure~\ref{fig2_inplane}. 
The currents shown in this figure %~\ref{fig2_inplane}
are directly proportional to the applied field but 
the slope of $j_x(B_y)$ is sample dependent.
The largest  slope is obtained for sample~A with 
the strongest asymmetry while the slope vanishes for the symmetric sample~E. 
In case of sample~D, having opposite SIA, also the slope is negative. 

%Primes should be introduced or a sentence that for 110 x,y,z have another meaning must be written

We consider MGE induced by radiation polarized along the $x$-axis, as used in this experiment. 
For this geometry, the MGE current  $j_x = \sum_{\beta}
\phi_{x \beta xx}\:B_\beta\:|E_x|^2$ is phenomenologically determined by the 
coupling of the $x$-component of the current polar vector with the 
axial vector of the magnetic field, because $|E_x|^2$ is an invariant in (110)-grown structure. 
Therefore, the photocurrent can occur only for certain 
$\bm{B}$-components which are transformed equally to $j_x$ for 
all symmetry operations. 
While the symmetry of perfectly symmetric (110)-grown QWs belongs 
to the point group  $C_{2v}$, asymmetric QWs belongs to the point group $C_s$. 
The point group $C_s$ contains only two symmetry elements:
identity and a mirror plane $m_1$, perpendicular to the $x$-axis. 
For $C_s$ group symmetry requirements are fulfilled for $j_x$ and $B_y$ or $B_z$ only. 
Indeed reflection by the  $m_1$-plane reverses the sign of 
$j_x$ ($j_x \rightarrow -j_x$) and two components of the magnetic field $\bm B$
($B_y \rightarrow -B_y$ and $B_z \rightarrow -B_z$). 
Thus the MGE can occur for  magnetic fields aligned in-plane 
and out-of-plane of the QW and the photocurrent
is given by
\begin{equation}\label{MPGE_Cs}
j_{x}^{\:C_{s}} 
= \phi^{SIA}_{xyxx}\:B_y\,|E_x|^2 + \phi^{BIA}_{xzxx}\:B_z\,|E_x|^2\:
\end{equation}
with tensors components $\phi^{SIA}_{xyxx}$ and $\phi^{BIA}_{xzxx}$
determined by the degree of the SIA and BIA, respectively. If  
the magnetic field is applied in $y$-direction 
the last term in equation~(\ref{MPGE_Cs}) becomes zero 
and the MGE current is determined solely by the degree of SIA. 

The experiment displayed in figure~\ref{fig2_inplane} 
%(samples~A to~E) 
shows that the magnitude of the $j(B_y)$-slope strongly 
depends on the  doping profile. Furthermore,
if the %sign of $\chi$ 
doping profile
is reversed (from sample~B to~D), the slope of the 
photocurrent gets reversed, too (see 
figure~\ref{fig2_inplane}).  
As the MGE current is proportional to the SIA coefficient, 
these observations demonstrate that the position of the doping layer
can be effectively used for tuning  the structure  
asymmetry strength. In particular,
the sign of  $\phi^{SIA}_{xyxx}$ can be inverted by putting the doping layer
%, keeping the distance fixed, 
to the other side of the QW.

\begin{figure}%[t]%[width=4cm]   % 12
\centerline{\epsfxsize 86mm \epsfbox{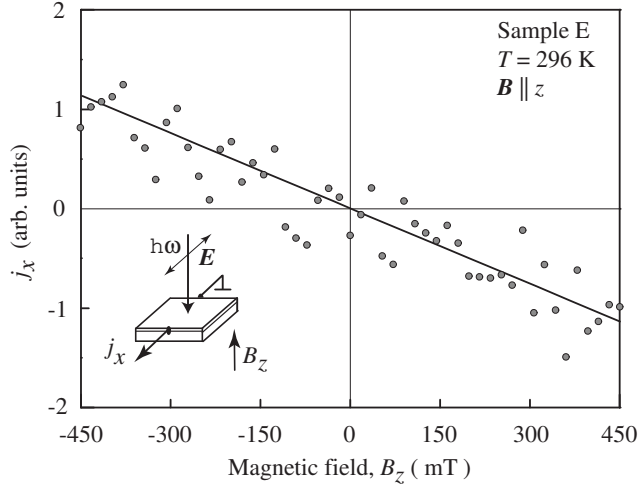}} % figure12  fig3_perpend
\caption{
%$J^{{\rm \: MPGE}}_x (B)$ for sample E measured for the radiation 
%polarized along $x$ and a magnetic field normal to the QWs.
%
Magnetic field dependence of $j_x$ 
for the radiation polarized along $x$-axis and 
an in-plane magnetic field normal to the QWs, $\bm{B} \parallel z$ (after  Ref.~\protect\cite{condmat110}).
}
\label{fig3_perpend}
\end{figure}

While for an in-plane magnetic field the photogalvanic effect, 
described by equation~(\ref{MPGE_Cs}), is observable in asymmetrical structures, 
it is forbidden in symmetrically 
grown QWs with the higher point group symmetry  $C_{2v}$.
This is caused by the presence of an additional mirror plane, $m_2$, being parallel to the 
QW plane of symmetrically grown (110)-structures. 
Indeed, reflection by this  plane does not modify $j_x$ 
but changes the polarity of in-plane axial vector $\bm B$. 
Therefore, in such systems linear coupling of $j_x$  and $B_y$ is forbidden.
On the other hand, mirror reflection of plane $m_2$ does not 
modify the $z$-component of $\bm B$. Thus the coupling of $j_x$ and 
$B_z$ is allowed for reflections by both, $m_1$ and $m_2$ planes,
and photocurrent $j_{x}$ can occur in symmetric (110)-oriented QWs
in the presence of a magnetic field in $z$-direction. 
For this symmetry equation~(\ref{phen0}) is reduced to
\begin{equation}\label{MPGE_C2v}
j_{x}^{\:C_{2v}} =  \phi^{BIA}_{xzxx}\:B_z\,|E_x|^2\:.
\end{equation}
This equation fully describes  the data taken from sample~E
obtained for in-plane magnetic field (figure~\ref{fig2_inplane}) and out-of-plane magnetic field (figure~\ref{fig3_perpend}). 
There, no MGE is observed for 
in-plane magnetic field but a sizeable effect is detected for 
$\bm{B}$ applied normal to the QW plane. In case of sample E, the absence 
of a magnetic field induced photocurrent 
in an in-plane $\bm{B}$  indicates that the QW is highly symmetric and 
lacks the structure  asymmetry.
The signal, observed in the same structure for an out-of-plane $B_z$-field 
stems from the BIA term (see equation~(\ref{MPGE_C2v})). 
Hence measurement of the MGE gives us an experimental handle
to analyze the degree of SIA.

\begin{figure}%[t]%[width=4cm]  % 13
\centerline{\epsfxsize 86mm \epsfbox{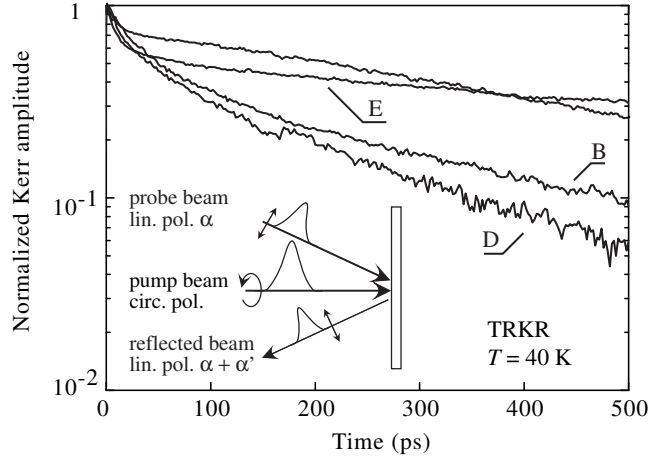}} % figure13  fig4_kerr
\caption{Kerr rotation measured in  
asymmetrically grown QWs samples~B, D and symmetrical structure~E.
We  note that the fast decaying components in the TRKR signals at 
short times are due to spin relaxation of photoexcited holes (after  Ref.~\protect\cite{condmat110}).
}
\label{fig4_kerr}
\end{figure}

The structural inversion asymmetry determines  the Rashba spin splitting 
and therefore controls the D'yakonov--Perel' 
relaxation~\cite{DP71} for spins aligned along the $z$-direction. Any
variation of SIA, e.g., due to asymmetric doping,  should result in a variation of 
the spin relaxation time. To directly demonstrate this connection, we compare spin relaxation rates measured 
in the symmetrically doped QW sample~E and the 
asymmetrically doped QW samples~B and~D (figure~\ref{fig4_kerr}). 
We extract the spin lifetime $\tau_s$  from time-resolved Kerr rotation (TRKR) measurements.
The time evolution of the Kerr rotation angle tracks the spin polarization
within the sample. By fitting an exponential decay function to the
data, $\tau_{s}$ is determined. In correspondence to the above  
photocurrent measurements, indicating 
a larger degree of asymmetry of the samples~B and~D compared to~E, 
%we observe that 
$\tau_{s}$ in sample~E
is found to be more than three times larger  than that in sample~B and about two
times larger than in sample~D. 

\section*{Acknowledgements} We thank E.~L.~Ivchenko, L.~E.~Golub and S.~A.~Tarasenko for helpful discussions. 
This work is supported by the DFG and RFBR.


\begin{thebibliography}{99}
\bibitem{footnoteklin} In the above analysis we do not consider the 
effect of a $\bm{k}$-linear spin splitting of the electron subband since 
it does not lead to a significant contribution to the pure spin current. 
In fact, to  first order in spin-orbit interaction, the spin splitting of 
the subband results only in a relative displacement of the spin-up and 
spin-down branches in $\bm{k}$-space. This shift does not disturb the 
symmetric distribution of carriers within each subband even if the system 
is driven into a non-equilibrium state, e.g. by absorption of linearly 
polarized radiation. Hence spin splitting in $\bm{k}$-space has no essential 
effect on the mechanism of spin-separation discussed here and can affect it 
only in high-order approximations
\end{thebibliography}
\end{document}